\documentclass[aps,prl,twocolumn,showpacs,superscriptaddress,letterpaper,amsmath,amssymb]{revtex4} 

\usepackage{graphicx}  
\usepackage{dcolumn}   
\usepackage{bm}        
\usepackage{amssymb}   
\usepackage{feynmf}
\usepackage{slashed}
\def\gsim{\lower0.5ex\hbox{$\:\buildrel >\over\sim\:$}}
\def\lsim{\lower0.5ex\hbox{$\:\buildrel <\over\sim\:$}}

\newcommand{\be}{\begin{equation}}
\newcommand{\ee}{\end{equation}}
\newcommand{\bea}{\begin{eqnarray}}
\newcommand{\eea}{\end{eqnarray}}

\newcommand{\nbox}{{\,\lower0.9pt\vbox{\hrule \hbox{\vrule height 0.2 cm
\hskip 0.2 cm \vrule height 0.2 cm}\hrule}\,}}

\begin{document}

\thispagestyle{empty}
\vspace*{-3.5cm}

\vspace{0.5in}

\title{Where are the Fermi Lines Coming From?}

\begin{center}
\begin{abstract}
We estimate the spatial locations of sources of the the observed features in the
Fermi-LAT photon spectrum at $E_\gamma=110$ and $E_\gamma=130$ GeV. We 
determine whether they are consistent with emission from a single source, as would
be expected in their interpretation as $\gamma\gamma$ and $\gamma Z$
lines from dark matter annhiliation, as well as whether they are
consistent with a dark matter halo positioned at the center of the
galaxy. We take advantage of the
per-photon measured incident angle  in
reconstructing the line features. In addition,  we use a data-driven background
model rather than making the assumption of a feature-less
background. We localize the sources of the features at 110
and 130 GeV. Assuming an Einasto (NFW) density model we find the
130 GeV line to be offset from the galactic center by 285 (280) pc, the 110 GeV line
by 60 (30) pc with a large relative separation of 220 (240)
pc. However, we find this displacement of each source from the galactic center,
as well as their relative displacement to be statistically consistent with a single Einasto or NFW dark matter halo at the center of the galaxy.
\end{abstract}
\end{center}

\author{Kanishka Rao}
\author{Daniel Whiteson}
\affiliation{Dept. of Physics \& Astronomy, UC Irvine, Irvine, CA }

\maketitle

\section{Introduction}

The search for the particle nature of dark matter is one of the
oustanding open questions of modern physics.  A broad program of
research with a variety of promising and complementary approaches has
become a major piece of experimental particle physics research. This includes searches for direct production at colliders,
searches for dark matter interactions with standard model particles in
large quiet underground detectors, and searches for signals of dark
matter annhiliation into standard model particles in regions of the
galaxy with large dark matter density.

A clear signal of dark matter annihilation may be carried by gamma
rays traveling to Earth from regions in the galaxy with high dark-matter
density.  As the photons do not typically scatter after their production,
the photon energy and direction are powerful handles for understanding the
mechanism of dark matter annihiliation into standard model particles.

One mechanism is annihilation resulting in quarks, which would
hadronize and yield $\pi^0$ particles which in turn produce
photons. The broad spectrum of such a process would give fairly low energy
photons ($E_\gamma \lesssim 50$ GeV) and may be
difficult to distinguish from other sources.

A clearer signature may appear from annihilation directly into
two-body final states including a photon, though the rate would be
smaller than continuum emission due to loop supression. Rather than yielding a broad
energy spectrum,  this process would
produce a photon with a well-defined
energy (a ``line'').  This makes a search for lines in the
photon spectrum an important component of the dark matter 
program using Fermi-LAT data~\cite{Ackermann:2011wa,Abdo:2010nc,Fermi:2012}.

Recently, observation of a feature with high local statistical
significance at $E_\gamma=130$ GeV was
reported~\cite{Weniger:2012tx,Tempel:2012ey}, with a source location
1-2 degrees away from the galactic center~\cite{finksu}. Follow-up analyses
suggested the possibility of two features, consistent with the spacing
of lines expected from $\gamma\gamma$ and $\gamma Z$
processes~\cite{twolines,wacker}

In this paper, we study in detail the question of the  location of the
source of the photons in the features. We confirm the location of the
photons in the feature at $E_\gamma=130$ GeV, and for the first time,  locate
the source of the photons in the feature at $E_\gamma=110$ GeV.  We study whether the
two are -- individually or collectively -- consistent with emission
from a  single source, as would be expected if they represent the
$\gamma\gamma$ and $\gamma Z$ processes, as well as whether that
source is consistent with a dark matter halo at the galactic center. In addition, we introduce a
new approach to analyzing the data which uses the
measured angles of the individual photons, rather than the median
photon angles.

\section{Energy Spectrum Analysis}

A source of photons with a very narrow range of energies would appear
in the spectrum of observed Fermi-LAT photons as a peak, due to finite
energy resolution.   Previous
analyses~\cite{Weniger:2012tx,twolines,wacker} have used the Fermi-LAT
energy dispersion formula~\cite{edisp} to build a probability density
function ({\emph{pdf}}) for the reconstructed energies of photons from a line,
and fit it to the observed spectrum.

\begin{figure}[h]
\begin{center}
\includegraphics[width=0.74\linewidth]{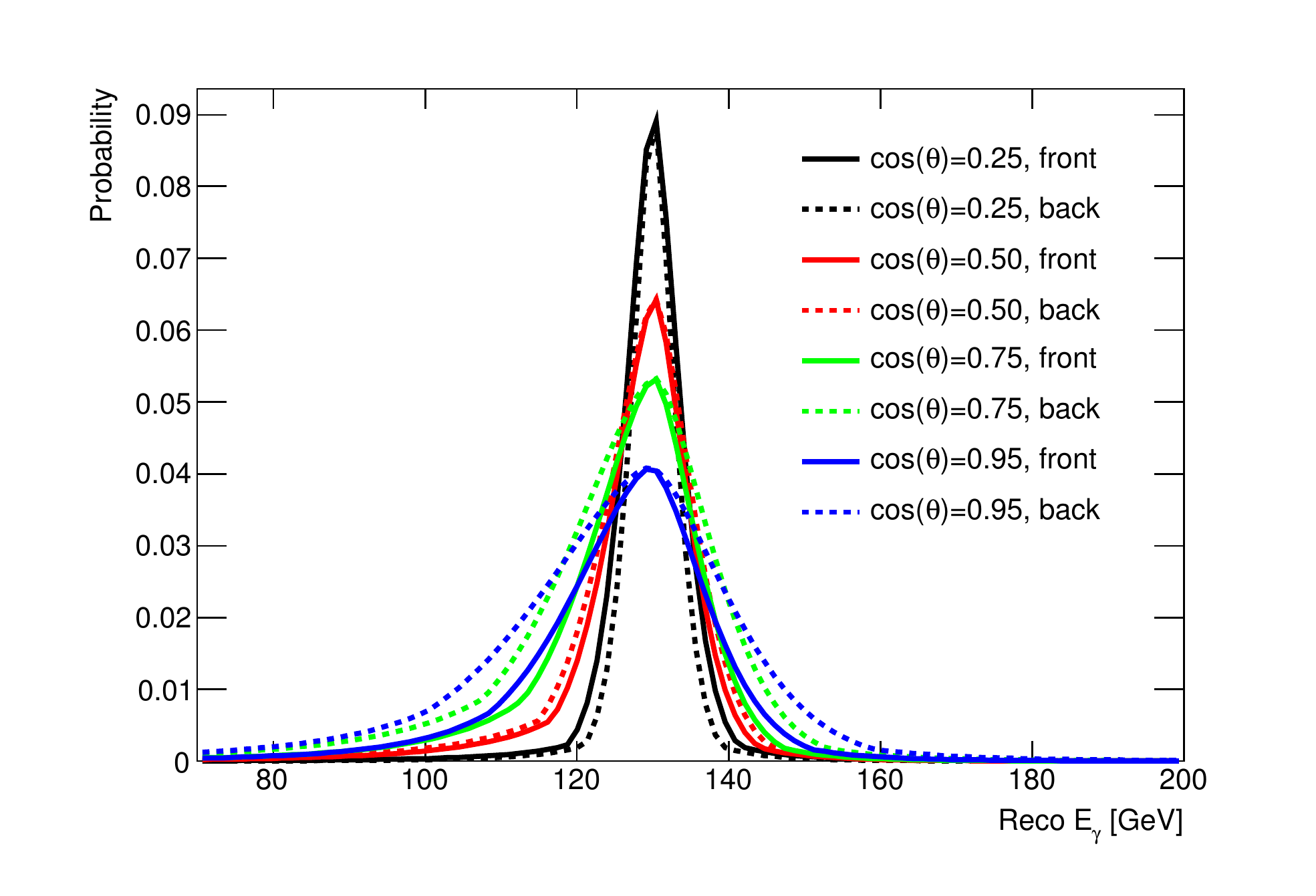}\\
\includegraphics[width=0.7\linewidth]{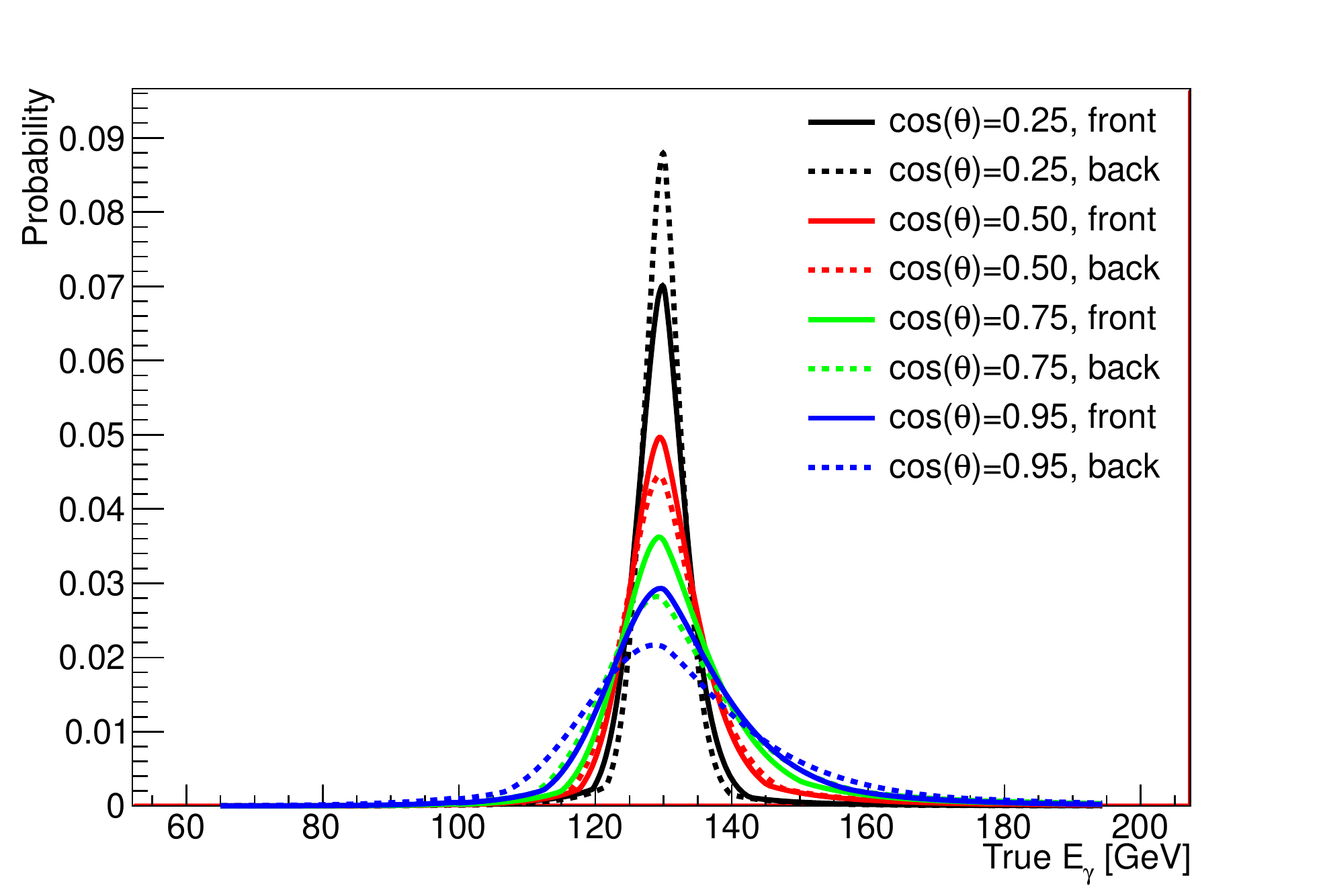}
\end{center}
\caption{Probability density functions which describe the energy
  dispersion of the Fermi-LAT.  Top, the probability density for
  reconstructed photon energies for a photon with
  $E_\gamma^{\textrm{true}}=130$ GeV.  Bottom,  the probability density for
  true photon energies for a photon with
  $E_\gamma^{\textrm{reco}}=130$ GeV. }
\label{fig:pdf}
\end{figure}

Given the small size of the dataset, we seek to use the maximum
available information about each photon to extract insight into the
observed features.   The energy resolution, for example, has a significant dependence on two other photon characteristics:
$\theta$, the incident angle relative to a line normal to the LAT
face, and the photon `type' which indicates whether the conversion happened in the front
or back layers of the tracker~\cite{defs}. Figure~\ref{fig:pdf} shows
the reconstructed energy pdf for various choices of $\theta$ and type.
This per-photon information is very pertinant to the analysis of this
spectrum. If the photons in the peak were all well-measured, it would
enhance the significance of the peak; conversely, if they were all
poorly-measured, it would degrade the significance.   

Our approach is to reconstruct the distribution of $E_\gamma^{\textrm{true}}$, the
true energy of photons striking the LAT, rather than the distribution
of $E_\gamma^{\textrm{reco}}$, the reconstructed energy of each
photon.  The former distribution can account for the per-photon
resolution, while the latter treats each photon equally.   The reconstructed energy is
related to the true energy by the pdf
$f(E^{\textrm{reco}}|E^{\textrm{true}},\theta,{\textrm{type}})$.   We use Bayes theorem to
invert this relation and calculate $f(E^{\textrm{true}}|E^{\textrm
  {reco}},\theta,{\textrm{type}})$ where $P(E^{\textrm{reco})}$ is taken from the full-sky photon  energy spectrum.  Other choices of priors have negliglible effect,  as the energy dispersion pdf is sharply peaked, see Fig~\ref{fig:pdf}.

The distribution of $E_\gamma^{\textrm{true}}$ is then built as a
kernel density estimate~\cite{kde} with
$f(E^{\textrm{true}}|E^{\textrm{reco}},\theta,{\textrm{type}})$ as the
choice of kernel.  Other approaches~\cite{Tempel:2012ey,finksu} have used adaptive
Gaussian kernels, but our physically motivated choice of kernel reflects the asymmetric
shape of the pdf and uses -- for the first time -- the per-photon
measured incident angle rather than the median angle of the dataset.

We use the publicly available Fermi-LAT data collected over 3.7
years. We use \texttt{ultraclean} photons with
zenith angle $<$ 100 to veto photons from Earth as well as standard
quality requirements~\cite{qual}. We select a
three-degree region around the galactic center.

\begin{figure}[h]
\begin{center}
\includegraphics[width=0.8\linewidth]{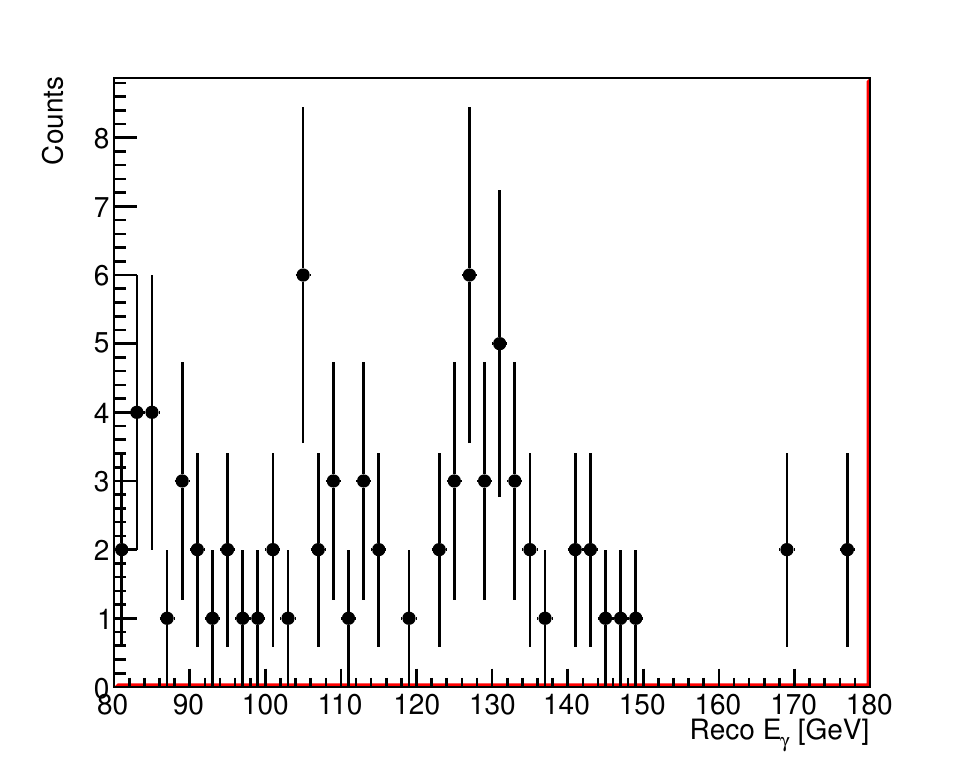}
\end{center}
\caption{ Reconstructed photon energies $E_\gamma^{\textrm{reco}}$ from Fermi-LAT in a
  three-degree region surrounding the galactic center. }
\label{fig:reco}
\end{figure}

\begin{figure}[h]
\begin{center}
\includegraphics[width=0.8\linewidth]{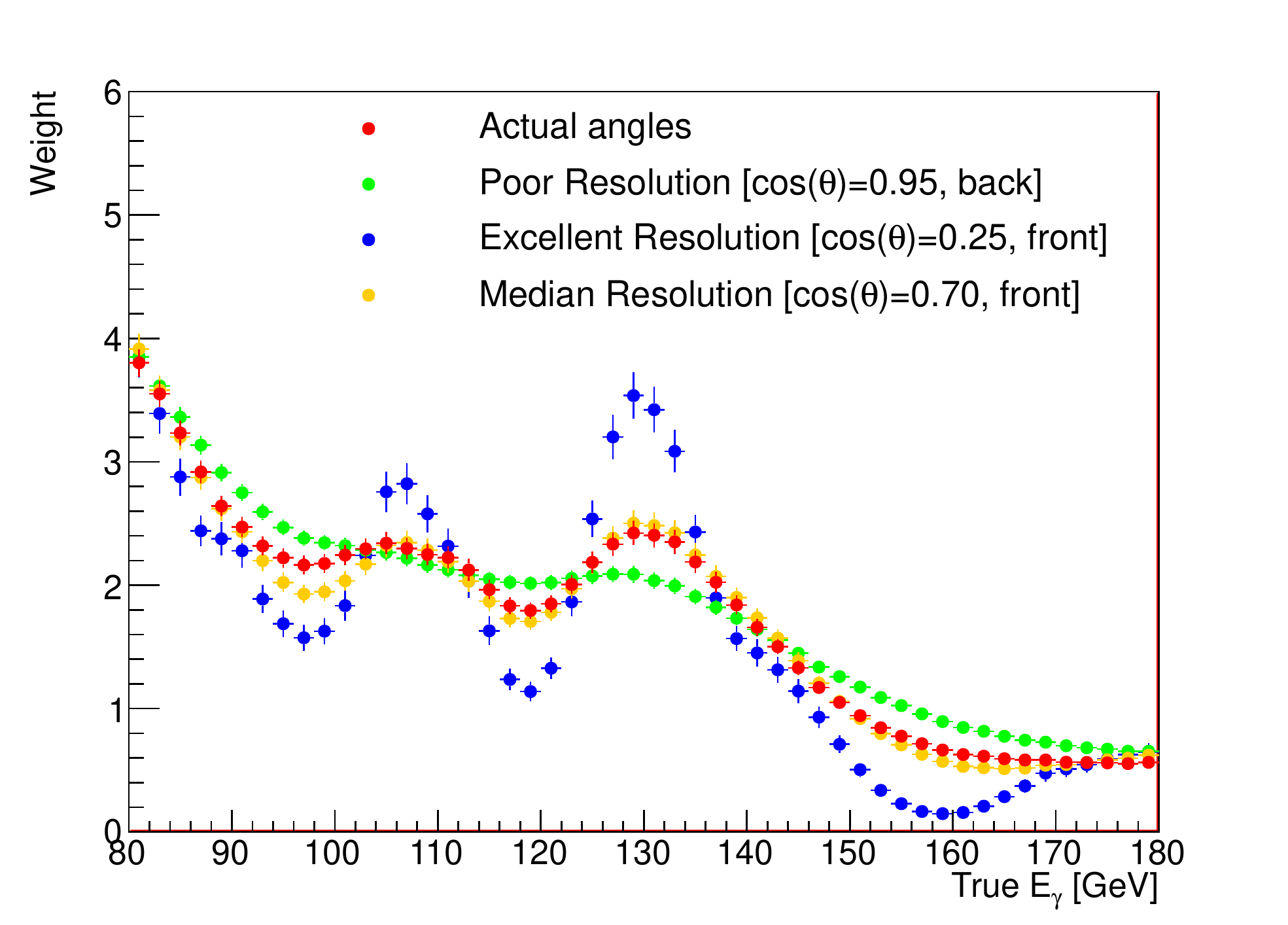}
\end{center}
\caption{ Relative likelihood of true photon energies,
  $E_\gamma^{\textrm{true}}$, built from the reconstructed energies of
  Fig~\ref{fig:reco} and the pdf
  $f(E^{\textrm{true}}|E^{\textrm{reco}},\theta,{\textrm{type}})$, see
  text for details. The impact of different assumptions about the
  photon incident angle $\theta$ is shown.}
\label{fig:reso}
\end{figure}

The observed photon counts in the three-degree region are shown in
Fig.~\ref{fig:reco}. The distribution of $E_\gamma^{\textrm{true}}$ is
shown in Fig.~\ref{fig:reso}, with several choices of angle-dependence
used to demonstrate the impact of this information.  Two choices
demonstrates the range of possibilities, if all photons had the
maximum or minimum possible energy resolutions. This demonstrates the
potential importance of this information. The spectrum which uses the
measured per-photon angles has peaks at 110 and 130 GeV which are
slightly degraded relative to a spectrum which uses the median photon angle.
 As the distribution of photons in the peak region and
the background region have similar distributions of $\cos(\theta)$ and
conversion type~\cite{splots}, there is not currently a large impact from using
the per-photon information, but it remains a useful method to
visualize the spectrum.

\section{Locating the Features}

\subsection{Technique}

We search for the most likely source position  using a
maximal likelihood fit of the observed photons to the
expected spatial and energy distributions from dark matter halos. The unbinned likelihood
is

\begin{equation} L(l,b)  = \prod_i^{N_i} \frac{
  n_{\textrm{bg}}f_{\textrm{bg}}(E_i,\theta_i) + n_{\textrm{sig}}f_{\textrm{sig}}(E_i,l_i,b_i | \theta_i,l,b )}{n_{\textrm{bg}}+n_{\textrm{sig}}}\end{equation}

\noindent
where $(l,b)$ are the galactic coordinates of the center of the dark
matter halo, $n_{\textrm{bg}}=$ and $n_{\textrm{sig}}$ are the
background and signal normalizations, the index $i$ runs over observed photon energy and
spatial parameters $(E_i,\theta_i,l_i,b_i)$ within the energy window
surrounding the feature(s) of interest, and $f_{\textrm{sig}}(E_i,l_i,b_i |\theta_i, l,b )$ is
the dark matter halo pdf as a function of photon energy $E_i$, incidence angle
$\theta_i$ at spatial position $(l_i,b_i)$ given a dark matter
halo centered at $(l,b)$. The pdf $f_{\textrm{sig}}$ accounts for per-photon
spatial~\cite{fermipsf} and energy~\cite{edisp} resolution in the same
spirit as the Bayesian unfolding described above; here we do not need
to explicitly unfold, as we have included the per-photon information
in the unbinned likelihood.

\begin{figure}[h]
\begin{center}
\includegraphics[width=0.8\linewidth]{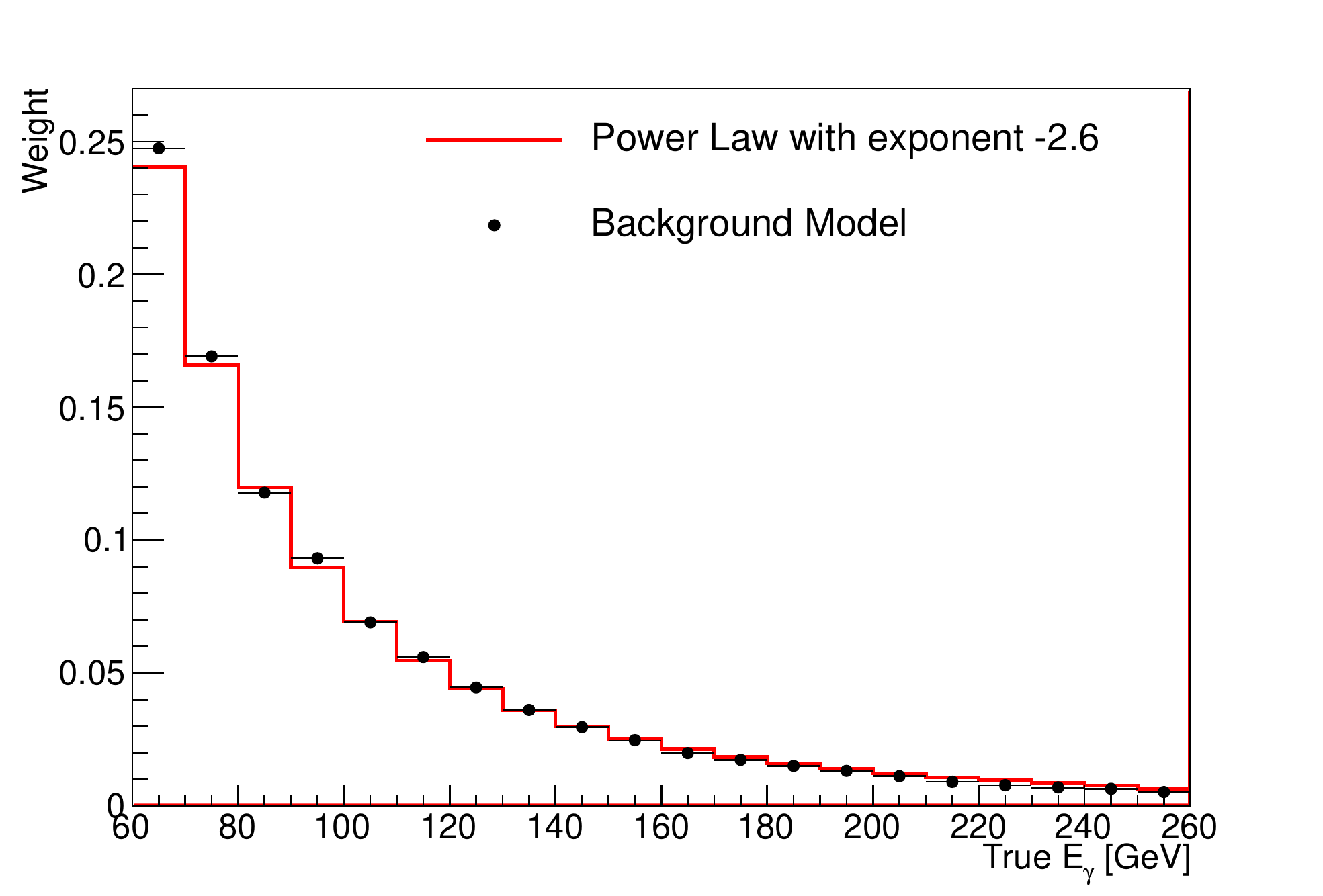}
\end{center}
\caption{The data-driven background model (points) constructed from
  photons outside the three-degree region surrounding the galactic
  center, compared to a power law with exponent -2.6.}
\label{fig:bg_form}
\end{figure}

The background pdf $f_{\textrm{bg}}$ is built from a data-driven model, with the energy dependence coming
from photons outside the three-degree region surrounding the galactic
center, see Figure~\ref{fig:bg_form}.  We find this to be reasonably
consistent with a power law model used in previous
analyses~\cite{Weniger:2012tx,Tempel:2012ey,finksu,twolines,wacker}. We use energy windows of $E_\gamma^{\textrm{true}} =
[105,115],[125,135]$ for the location of the feature at 110
GeV and 130 GeV, respectively, or both windows for the combined features. In the case of
the two-feature analysis ($E_\gamma=110,130$ GeV), there are two $f_{\textrm{sig}}$ terms in
the likelihood, one for each peak.  We use normalizations of
$n_{\textrm{sig}}=6$ and $n_{\textrm{sig}}=14$ for the $E_\gamma=110$
and $130$ GeV features, respectively.

The dark matter halo pdfs $f_{\textrm{sig}}$ are derived from either NFW~\cite{nfw} or
Einasto~\cite{einasto} halo profiles; the pdf in $(l,b$) is calculated
via the line-integral of the square of the dark matter density~\cite{130loc}.  We use $\alpha_{E}$ = 0.17 for
the Einasto model and $r_{s}$ = 20 kpc for both models.  

\subsection{Results}

The position
of the most likely values of the halo centers are given in
Table~\ref{tab:pos} for both DM profiles and each of the features as
well as
the combined spectrum.  In both cases, the feature at 130 GeV appears
to be displaced from the galactic center, as previously reported, but
the 110 GeV feature appears to be centered at the galactic origin.

\begin{figure}[h]
\begin{center}
\includegraphics[width=0.9\linewidth]{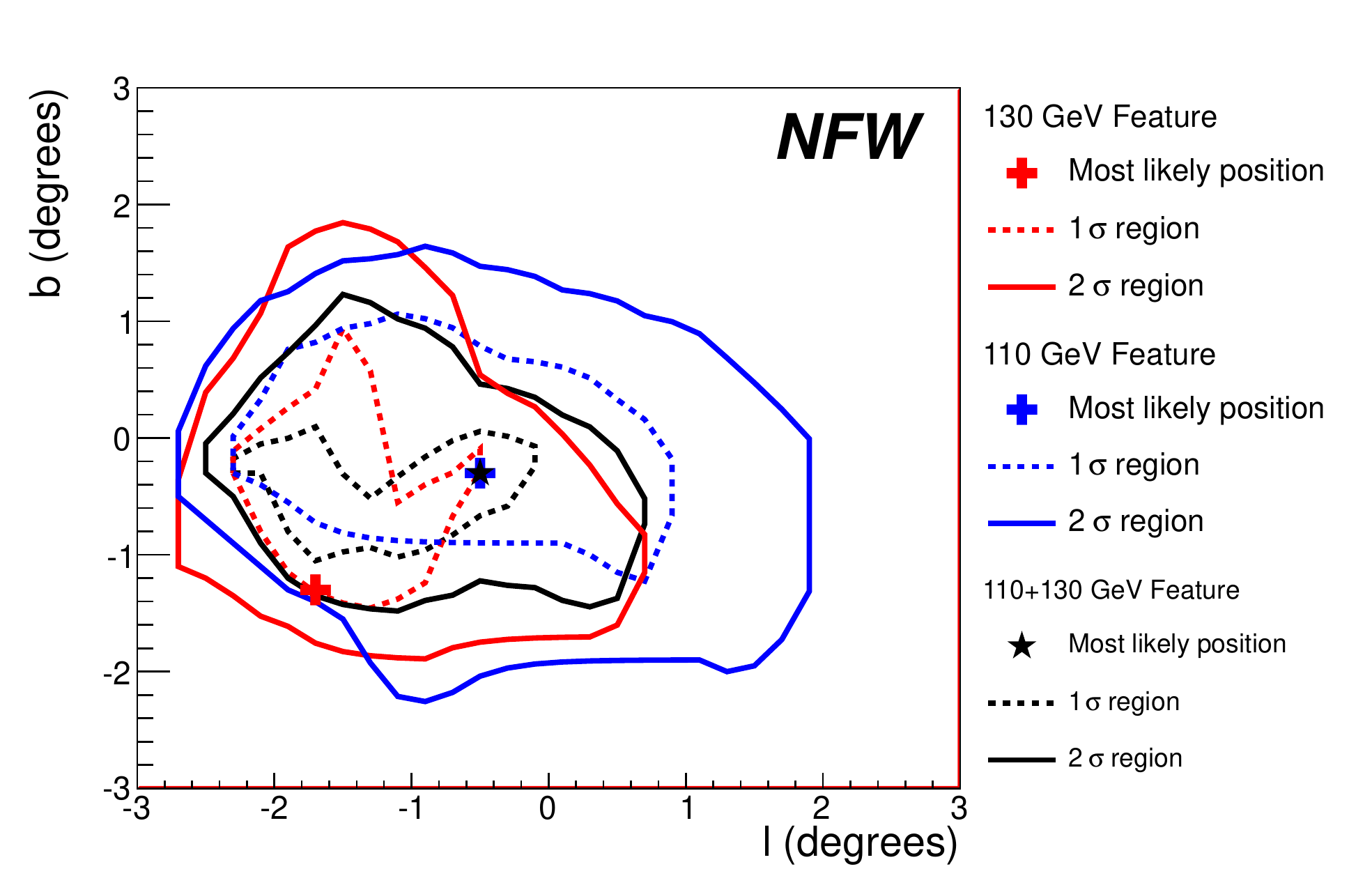}
\includegraphics[width=0.9\linewidth]{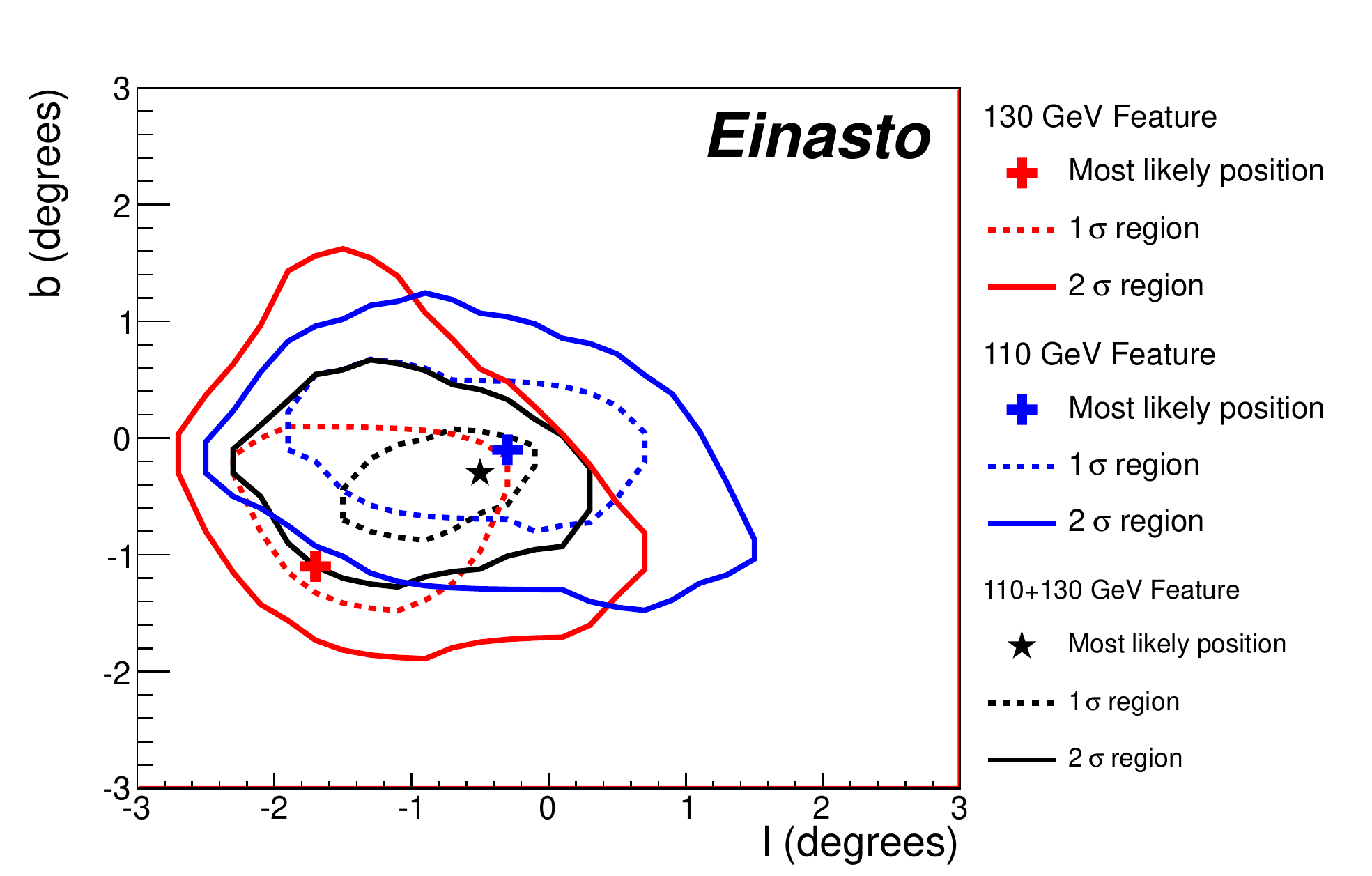}
\end{center}
\caption{Most likely positions in galactic coordinates $(l,b)$ of an
  NFW (top) or Einasto (botton) dark matter halo, maximized seperately for the
  features at 130 GeV, 110 GeV, or the combined features. Also shown
  are $1\sigma$ and $2\sigma$ uncertainty regions, see text for details.}
\label{fig:maps}
\end{figure}

Figure~\ref{fig:maps} shows the most likely positions for each of the
dark matter halo profiles and each of the energy spectrum features, as well as uncertainty regions. The
uncertainty regions are calculated in a frequentist manner, using
simulated experiments to determine the $\Delta L$ threshould which
would contain the true $(l,b)$ position in 68\% (for $1\sigma$) or 95\%
(for $2\sigma)$ of the cases.  

\begin{figure}[ht]
\begin{center}
\begin{minipage}[b]{0.49\linewidth}
\includegraphics[width=\linewidth]{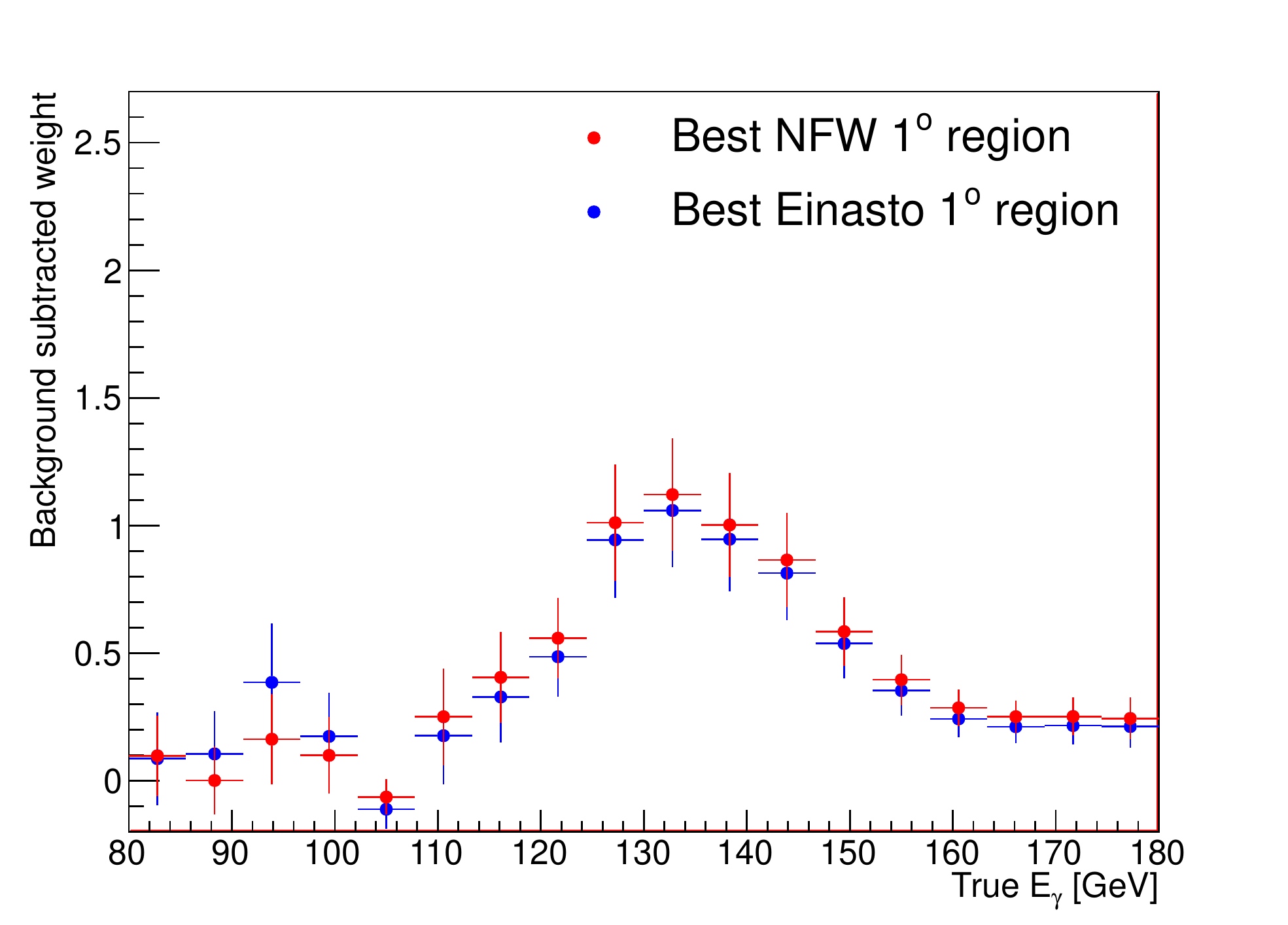}
\end{minipage}
\begin{minipage}[b]{0.49\linewidth}
\includegraphics[width=\linewidth]{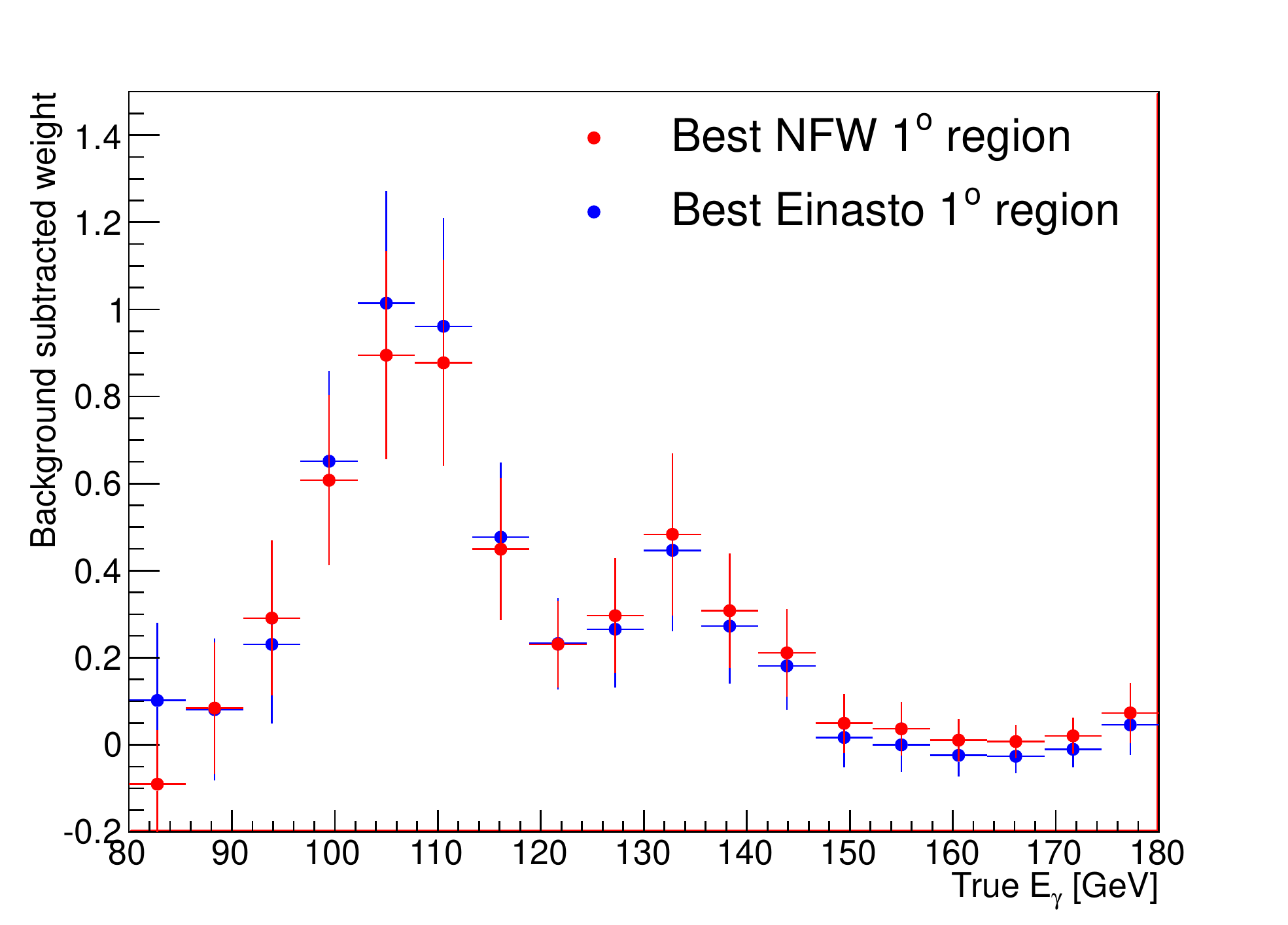}
\end{minipage}
\end{center}
\caption{ Spectrum in a region within 1 degree of the maximum
  likelihood for 130 (left) and 110 (right) features assuming an
  Einasto or NFW source.}
\label{fig:peaks1deg}
\end{figure}

\begin{figure}[ht]
\begin{center}
\begin{minipage}[b]{0.49\linewidth}
\includegraphics[width=\linewidth]{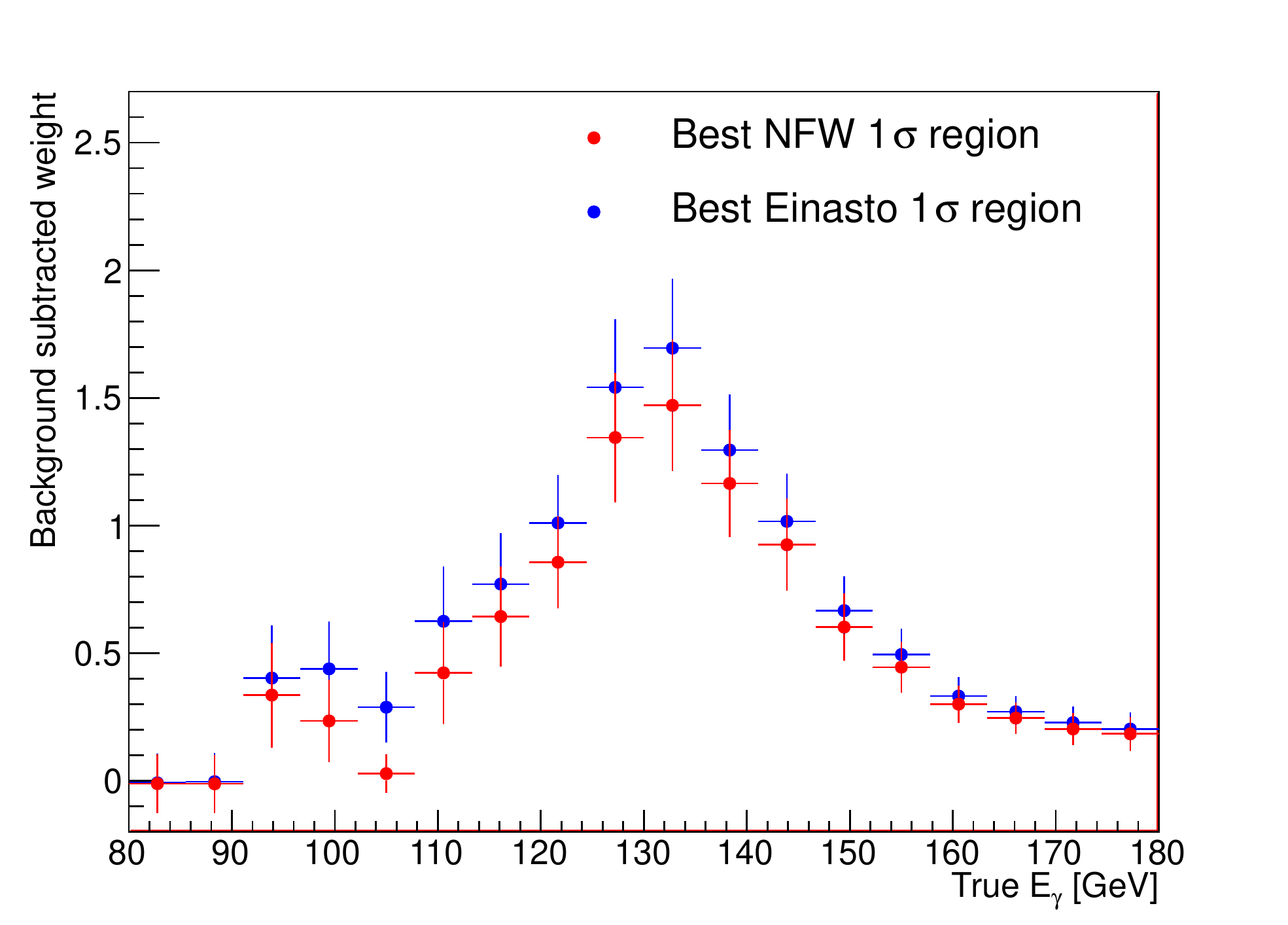}
\end{minipage}
\begin{minipage}[b]{0.49\linewidth}
\includegraphics[width=\linewidth]{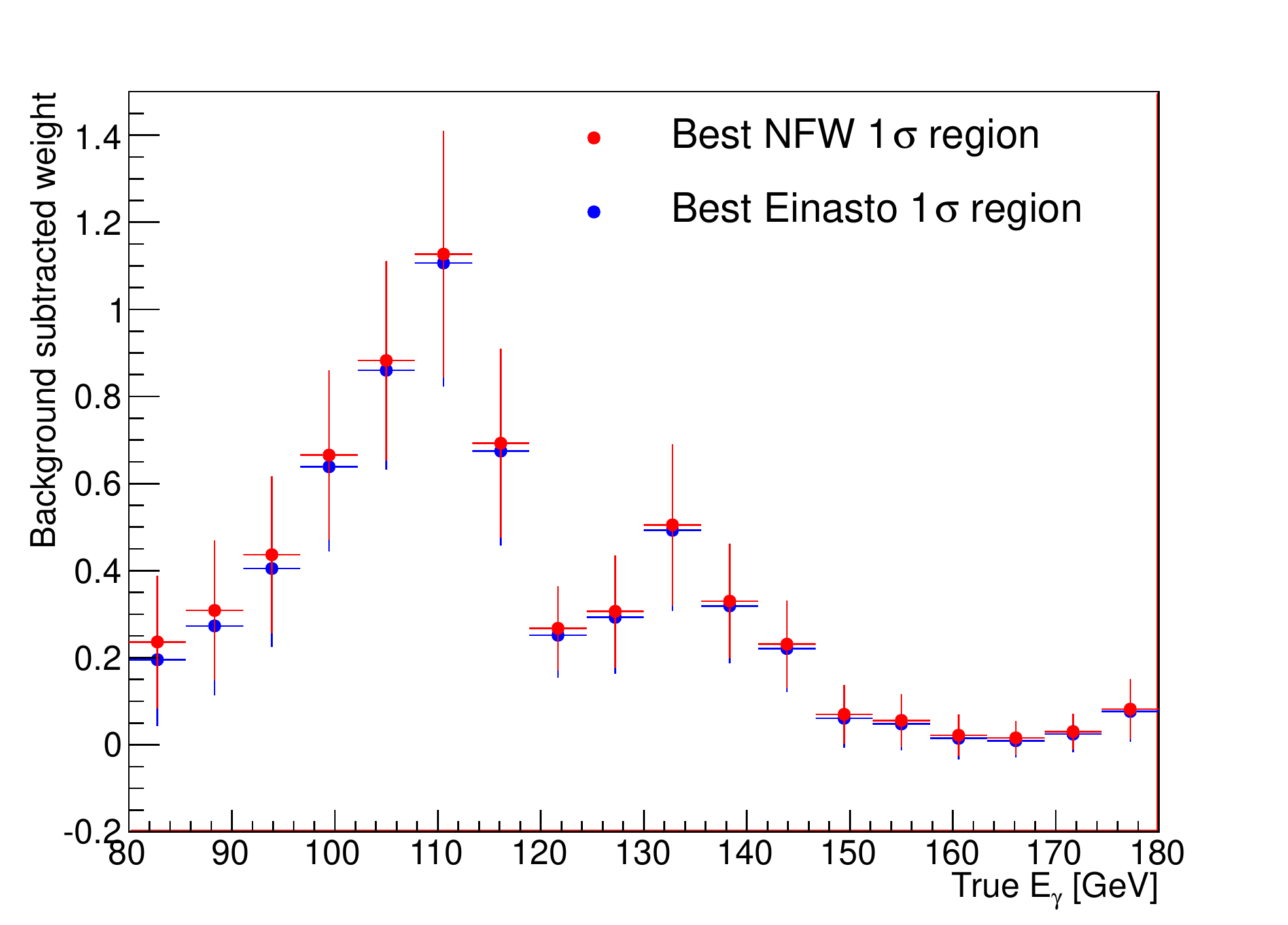}
\end{minipage}
\end{center}
\caption{ Spectrum in a region within 1$\sigma$ of the maximum
  likelihood for 130 (left) and 110 (right) features assuming an
  Einasto or NFW source.}
\label{fig:peaks1sigma}
\end{figure}

 If the two features were both due to dark matter annhililation processes,
one would expect them to be co-located. Figure~\ref{fig:peaks1deg}
shows the unfolded $E^{\textrm{true}}_\gamma$ spectrum in a 1-degree circle surrounding each of the
features; in the spectrum which highlights the 130 GeV feature, the
110 GeV feature is suppressed, and vice versa.  Similarly,  Figure~\ref{fig:peaks1sigma} shows
the unfolded $E^{\textrm{true}}_\gamma$ spectrum in the $1\sigma$ region surrounding each feature.

\begin{table} 
\caption{Most likely positions in galactic coordinates $(l,b)$ of an
  NFW or Einasto dark matter halo, maximized seperately for the
  features at 130 GeV, 110 GeV, or the combined features.}
\label{tab:pos}
\begin{tabular}{ l c c c}
\hline\hline
       &130 GeV&110 GeV& Combined \\ \hline
NFW & (-1.5,-1.1) & (-0.2,0.1) & (-0.4,-0.2)\\
Einasto &(-1.5,-1.2) & (-0.4,-0.2) & (-0.2,0.1)\\
\hline\hline
\end{tabular}
\end{table}

This suggests that the two features of the energy spectrum may have spatially distinct sources,
and these sources may themselves be offset from the galactic center.
To evaluate the significance of the sources' displacement from the
galacter center and from each other a more rigorous analysis which accounts for uncertainties
is performed below.

\subsection{Interpretation}

If the observed features are both due to dark matter annhilation, then
their locations in the sky depend only on the dark matter density and
not the nature of the final state, $\gamma\gamma$ or $\gamma Z$, and
so should yield locations consistent with a single source. As shown
above, we find that the sources of the $E_\gamma=110$ and 130 GeV features
are seperated appreciably from each other, and that the $E_\gamma=130$ feature is
displaced from the galactic center.

Definitive conclusions, however, require analysis of the uncertainties
involved, including the expected fluctuations of the measured source
locations given the small number of photons in each feature. We seek to answer two questions:
\begin{itemize}
\item Are the two features consistent with emission from a single
  source?
\item Are the features consistent with emission from a dark matter
  halo at the galactic center?
\end{itemize}

\begin{figure}[h]
\begin{center}
\begin{picture}(80,90)
\put(0,0){\includegraphics[width=0.9\linewidth]{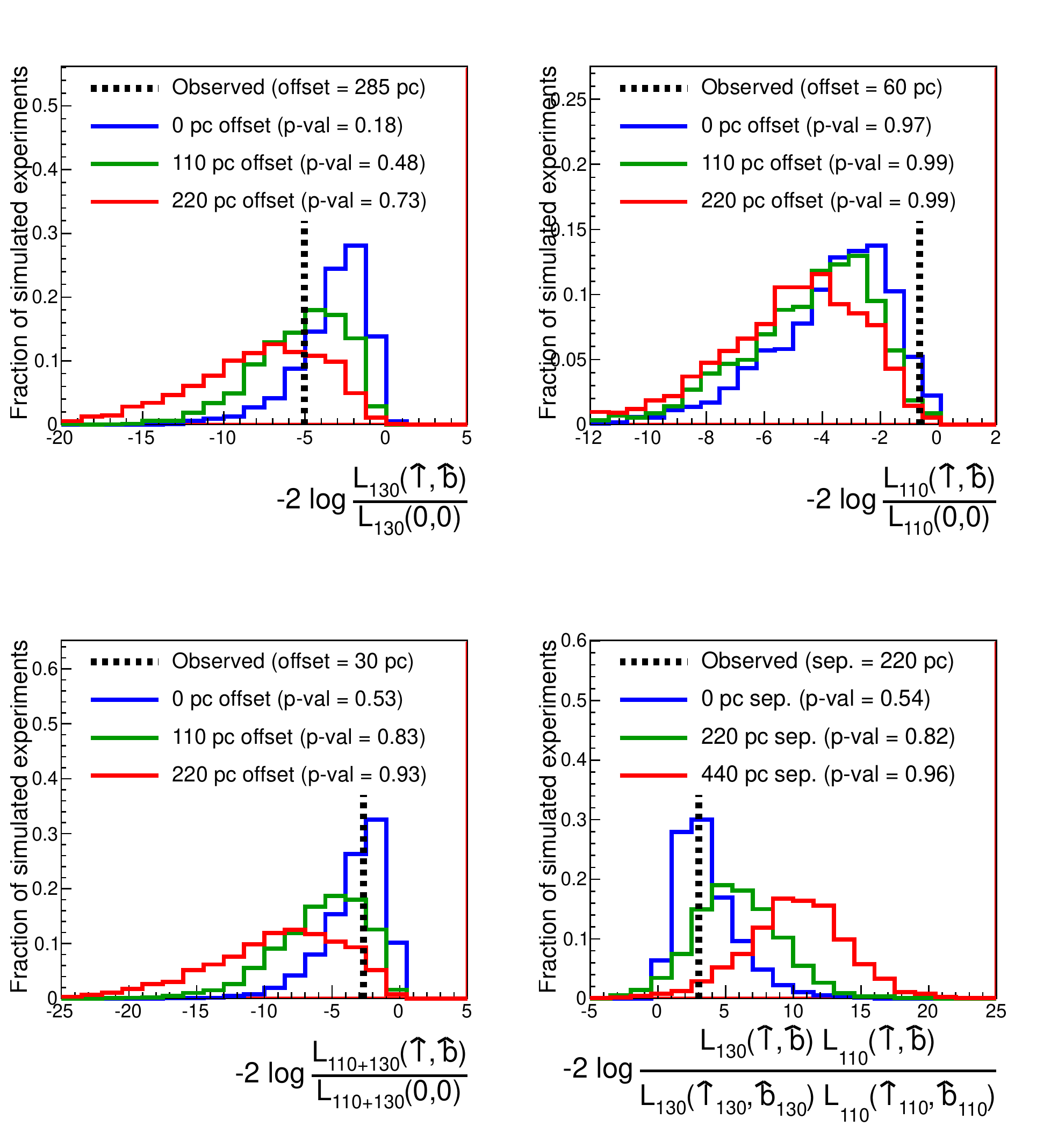}}
\put(6,60){(a)}
\put(45,60){(b)}
\put(6,15){(c)}
\put(68,15){(d)}
\end{picture}
\end{center}
\caption{ The log of likelihood ratios comparing the likelihood of a Einasto
  DM halo placed at the most likely position ($\hat{l},\hat{b}$) or at
  the galactic center $(l=0,b=0)$, in the case of the 130 GeV feature
  (a), the 110 GeV feature (b), or the combined features (c). Also
  shown is the likelihood ratio between two co-located sources or
  sources which are seperated (d). In each case, we show the expected
  distribution in simulated experiments.}
\label{fig:llr_ein}
\end{figure}
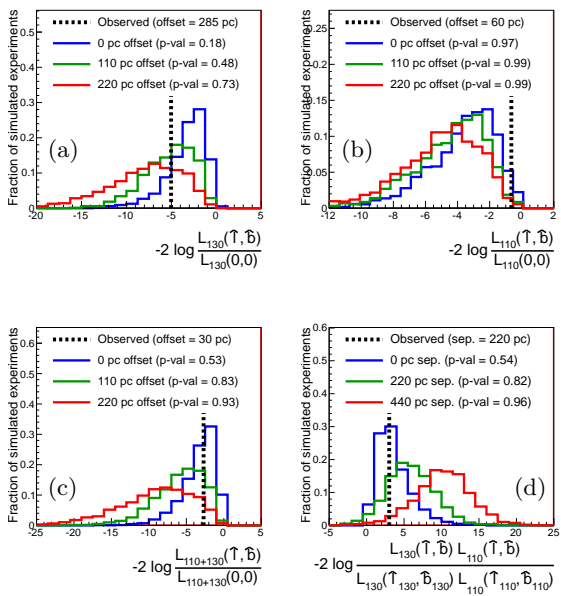

\begin{figure}[h]
\begin{center}
\begin{picture}(80,90)
\put(0,0){\includegraphics[width=0.9\linewidth]{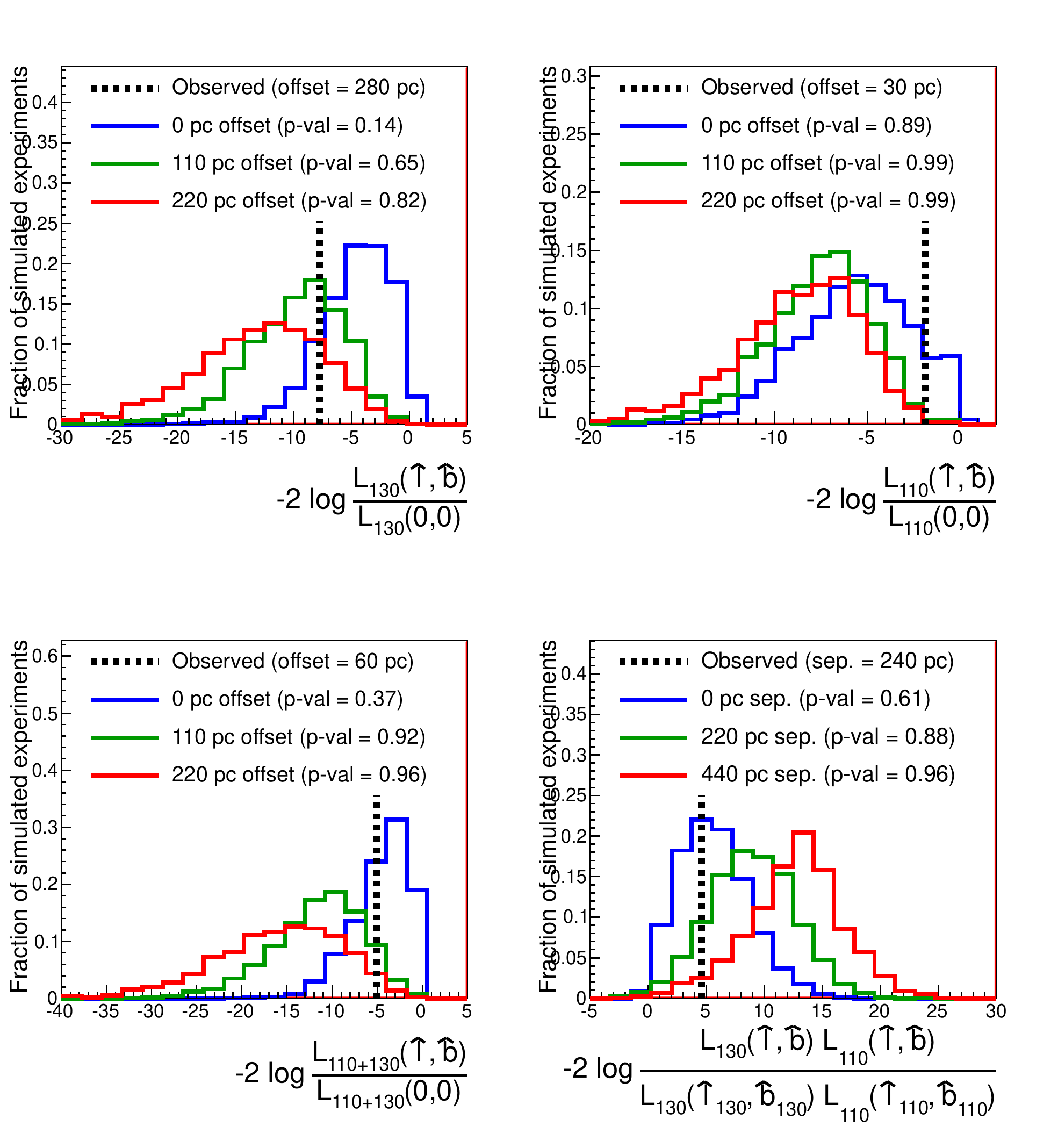}}
\put(6,60){(a)}
\put(45,60){(b)}
\put(6,15){(c)}
\put(68,15){(d)}
\end{picture}
\end{center}
\caption{ The log of likelihood ratios comparing the likelihood of an NFW
  DM halo placed at the most likely position ($\hat{l},\hat{b}$) or at
  the galactic center $(l=0,b=0)$, in the case of the 130 GeV feature
  (a), the 110 GeV feature (b), or the combined features (c). Also
  shown is the likelihood ratio between two co-located sources or
  sources which are seperated (d). In each case, we show the expected
  distribution in simulated experiments.}
\label{fig:llr_nfw}
\end{figure}
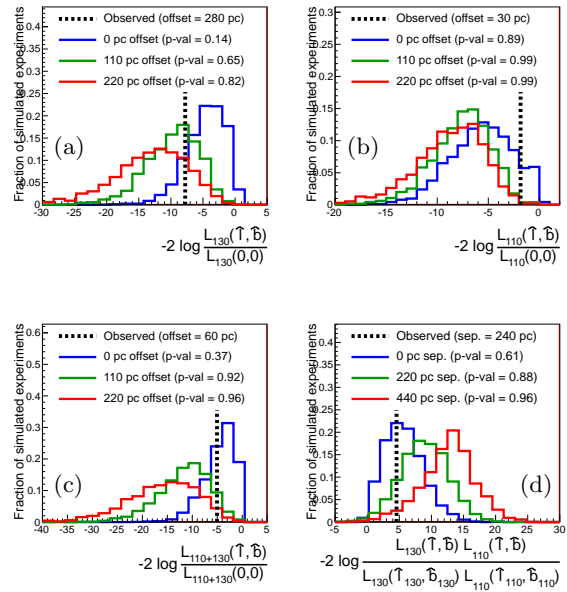

We answer these questions by comparing hypotheses using likelihood
ratios. For example, we can probe whether the feature at $E_\gamma=
130$ GeV is
consistent with emission from the galactic center by evaluating

\begin{equation} q = - 2 \log\frac{L(l=\hat{l},b=\hat{b})}{L(l=0,b=0)} \end{equation}

\noindent
where $(\hat{l},\hat{b})$ is the location which maximizes $L$ and
$(0,0)$ serves as the null hypothesis.   When
$(\hat{l},\hat{b})$ is close to $(0,0)$, the log of the likelihood
ratio $q$ approaches zero; when it is far from the origin, it becomes
negative.   We compare the measured value of $q$ to the expected
distribution in simulated experiments from an NFW or Einasto profile
with a line at $E_\gamma=130$ GeV placed at (0,0), see Figures~\ref{fig:llr_ein}(a) and
~\ref{fig:llr_nfw}(a).  While the value of $q$ indicates that the
maximal position is displaced from the origin, it is consistent
 with
the expected distribution of $q$ values when the DM halo is at the
origin; the $p$-values = 0.18 and 0.14 for Einasto or NFW, respectively.
 Distributions of $q$ for DM halos offset from the origin are
also shown.

We perform this analysis for the feature at $E_\gamma=110$ GeV as well, see
Figures~\ref{fig:llr_ein}(b) and ~\ref{fig:llr_nfw}(b).  As this
feature appears to be located close to $(0,0)$, the $q$ values are
close to zero and entirely consistent with a DM halo at the galactic
center.

We repeat the analysis for the combined features at 110 and 130 GeV,
where we constrain them to have the same location, see
Figures~\ref{fig:llr_ein}(c) and ~\ref{fig:llr_nfw}(c). Again, the $q$
values are consistent with a DM halo at the galactic center.

To address the question of whether the two features are more likely to
be from a single source (consistent with a DM interpretation) or from
two spatially distinct sources, we construct a likelihood ratio

\begin{equation} q = - 2 \log\frac{L_{130}(\hat{l},\hat{b}) \times
  L_{110}(\hat{l},\hat{b})
}{L_{130}(\hat{l}_{130},\hat{b}_{130}) \times
  L_{110}(\hat{l}_{110},\hat{b}_{110})} \end{equation}

\noindent
where in one case we find the single location $(\hat{l},\hat{b})$
which maximizes $L$ for both features, and in the other we allow the
two features to find individual maxima:
$(\hat{l}_{110},\hat{b}_{110})$ and $(\hat{l}_{130},\hat{b}_{130})$.
This $q$ will be zero if the individually optimized (and possibly separated) locations are
consistent with the most likely single joint location, and positive
if individually optimized  (and possibly separated) locations give a better description of the
data.  Again we compare to the expected distribution of the likelihood
ratio in simulated experiments in which the two sources are co-located
and find that the observed $q$ is consistent with a co-located source, see Figures~\ref{fig:llr_ein}(d) and ~\ref{fig:llr_nfw}(d).

\section{Conclusion}

We haved demonstrated via Bayesian unfolding the
potential importance of including the per-photon energy resolution,
though for this dataset the results are little different from the
standard approach of assuming the median angle and therefore resolution.

We have presented an analysis of the spatial locations of the sources
of the observed features in the Fermi-LAT energy spectrum at
$E_\gamma=110$ and 130 GeV.  We find:

\begin{itemize}
\item  the 130 GeV photons come from a location
  displaced from the galactic center, but consistent with an Einasto or NFW dark matter halo at the
  center. This is consistent with previous work~\cite{130loc}.
\item  the 110 GeV photons come from a location
  close to the galactic center, and consistent with an Einasto or NFW dark matter halo at the center.
\item  the two features at 110 and 130 GeV photons are consistent with a single
  DM halo at the galactic center, with either an Einasto or NFW density profile.
\end{itemize}

Some have suggested that the displacement of the source of the
$E_\gamma=130$ GeV feature indicates that DM halos may be displaced from the galactic
center~\cite{kuhlen}, but we find that the current data are consistent
with a DM halo from the center.


 
\section{Acknowledgements}

The authors acknowledges  contributions, explanations
and useful discussions with Eric Albin which are clearly deserving of
authorship, clever statistical insight from Kyle Cranmer, useful comments from Annika Peter, and technical support from Mariangela Lisanti and Tracy Slatyer.
DW and KR are supported by grants from the Department of Energy
Office of Science and by the Alfred P. Sloan Foundation. DW is
grateful to the Aspen Center for Physics, where a portion of this work was
performed and supported by NSF grant no. 1066293.


\begin{thebibliography}{99}



\bibitem{Ackermann:2011wa}
  M.~Ackermann {\it et al.}  [Fermi-LAT Collaboration],
  Phys.\ Rev.\ Lett.\  {\bf 107}, 241302 (2011)
  [arXiv:1108.3546 [astro-ph.HE]];

\bibitem{Abdo:2010nc}
  A.~A.~Abdo {\it et al.}  [ The Fermi-LAT Collaboration],
  Phys.\ Rev.\ Lett.\  {\bf 104}, 091302 (2010)
  [arXiv:1001.4836 [astro-ph.HE]].

\bibitem{Fermi:2012}
  M.~Ackermann {\it et al.}  [Fermi-LAT Collaboration],
  arXiv:1205.2739 [astro-ph.HE].



\bibitem{Weniger:2012tx}
  C.~Weniger,
  arXiv:1204.2797 [hep-ph].

\bibitem{Tempel:2012ey}
  E.~Tempel, A.~Hektor and M.~Raidal,
  arXiv:1205.1045 [hep-ph].

\bibitem{finksu}
M.~Su and D.~P.~Finkbeiner, (2012),
  arXiv:1206.1616 

\bibitem{twolines}
  A.~Rajaraman, T.~M.~P.~Tait and D.~Whiteson, JCAP, accepted (2012),
 JCAP {\bf 1209}, 003 (2012)
 [arXiv:1205.4723 [hep-ph]].

\bibitem{wacker}
 T.~Cohen, M.~Lisanti, T.~R.~Slatyer and J.~G.~Wacker,
  arXiv:1207.0800 [hep-ph].

\bibitem{edisp}
\url{http://fermi.gsfc.nasa.gov/ssc/data/analysis/documentation/Cicerone/Cicerone_LAT_IRFs/IRF_E_dispersion.html}

\bibitem{defs}
\url{http://fermi.gsfc.nasa.gov/ssc/data/analysis/documentation/Cicerone/Cicerone_Data/LAT_Data_Columns.html}

\bibitem{kde} L. Holmstr\"om, H. Miettinen, and S. R. Sain, Comp. Phys. Comm. 88 (1995) 195-210.



\bibitem{qual} Pass7,  \texttt{ultraclean} class, quality
  requirements: {\mbox{\texttt{DATA\_QUAL$=1$ \&\& LAT\_CONFIG$=$1 \&\& ABS(ROCK\_ANGLE)$\le$52}}}

\bibitem{splots}
 D.~Whiteson,
  arXiv:1208.3677 [astro-ph.HE].

\bibitem{fermipsf}
  \url{http://fermi.gsfc.nasa.gov/ssc/data/analysis/documentation/Cicerone/Cicerone_LAT_IRFs/IRF_PSF.html}
    

\bibitem{nfw} 
J.~F.~Navarro, C.~S.~Frenk and S.~D.~M.~White,
  Astrophys.\ J.\  {\bf 462}, 563 (1996)
  [astro-ph/9508025].

\bibitem{einasto} 
 J. Einasto and U. Haud , Astron. Astrophys. 223, 89 (1989);
 A.~W.~Graham, D.~Merritt, B.~Moore, J.~Diemand and B.~Terzic,
  Astron.\ J.\  {\bf 132}, 2685 (2006).

\bibitem{130loc}
  R.~-Z.~Yang, Q.~Yuan, L.~Feng, Y.~-Z.~Fan and J.~Chang,
  Phys.\ Lett.\ B {\bf 715}, 285 (2012)

\bibitem{kuhlen}
 M.~Kuhlen, J.~Guedes, A.~Pillepich, P.~Madau and L.~Mayer,
  arXiv:1208.4844 [astro-ph.GA].

\end{thebibliography}
\end{document}